\documentclass[
preprint,
prd,
aps,
amsmath,
amssymb,
]{revtex4-1}

\usepackage{graphicx}
\usepackage{color}
\usepackage{float}
\usepackage{subfloat}
\usepackage{subfigure}

\usepackage{hyperref}

\begin{document}
\title{All about $H^{\pm\pm}$ in Higgs Triplet Model}
\author{Ambalika Biswas}\email{ani73biswas@gmail.com\,,\,ambalika12t@boson.bose.res.in}
\affiliation{S. N. Bose National Centre For Basic Sciences,\\
 Block JD, Sector III, Salt Lake, Kolkata 700106, INDIA}
\date{\today}

\begin{abstract}
  In this piece of work, the nonstandard Higgs boson  masses have been constrained by restrictions arising from Higgs diphoton decay width, stability, perturbativity, unitarity and electroweak T-parameter in the framework  of the Higgs triplet model. As a matter of  fact the loop  induced Higgs decays will  now obtain additional contributions from singly and doubly charged Higgs bosons. In simulations of pair production of $H^{++}$ it is assumed that the production channel $q\overline{q}\rightarrow \gamma^{*},Z^{*}\rightarrow H^{++}H^{--}$ is the only mechanism. Whereas there may be other channels of production of doubly charged Higgs bosons as for instance from  the decay of singly charged Higgs bosons. The effect of this additional sources of production of the contributing non-standard charged particles to the  loop induced Higgs decays has been studied in this paper.   
\end{abstract}

\maketitle

\section{Introduction}
The Higgs Triplet Model abbreviated as HTM~\cite{Konetschny:1977, Mohapatra:1980, Magg:1980, Schechter:1980, Cheng:1980} has been extensively studied in relation to neutrino mass generation with a non-minimal Higgs sector. In Higgs triplet models the SM Lagrangian is augmented solely by $\Delta$ which is a $SU(2)_L$ triplet of scalar particles with hypercharge $Y=2$. Neutrinos may obtain mass via the vacuum expectation value (vev) of a neutral Higgs boson in an isospin triplet representation. The smallness of neutrino masses is guaranteed by the smallness of the triplet vev which is assumed to be less than 1 GeV and the non-conservation of lepton number which is explicitly broken in the scalar potential of the Higgs triplet model by a trilinear coupling $\mu$. The trilinear coupling $\mu$ is protected by symmetry and is naturally small which assures the smallness of neutrino masses. 

The model predicts several scalar particles, including a doubly charged Higgs boson ($H^{\pm\pm}$) and a singly charged Higgs boson ($H^{\pm}$), for which direct searches
are being carried out at the LHC~\cite{CMS:2011, ATLAS:2012}. Apart from these there is a CP odd neutral scalar, $A^{0}$ and two CP even neutral scalars, H and h. The rest degrees of freedom are absorbed by the vector bosons. In a large part of the parameter space of the HTM the lightest CP-even scalar,h, has essentially the same couplings to the fermions and vector bosons as the Higgs boson of the SM~\cite{Dey:2009, Akeroyd:2010, Arhrib:2011}. Now coming to the loop induced Higgs decay, $h\rightarrow \gamma \gamma$ would receive additional contributions from the non-standard charged Higgs bosons both doubly and singly charged ones. Thus this additional contribution  may result in a decay width which matches the present LHC decay width results~\cite{PDG:2014, ATLASCMS:2015}. $\lambda_{1}$ which is a quartic coupling in the potential of the Higgs triplet model to be discussed very soon, controls the contribution of $H^{\pm\pm}$ to the Higgs diphoton decay width. The case of $\lambda_{1}>0$ leads to destructive interference between the combined SM contribution (from W and fermion loops) and the contribution from $H^{\pm\pm}$ as was studied in~\cite{Arhrib:2012}. Later on the case for $\lambda_{1}<0$ was studied in~\cite{Akeroyd:2012} which leads to constructive interference.

In such a situation of constructive interference the production of $H^{\pm\pm}$ draws a great deal of attention. Production channels, $q\overline{q}\rightarrow \gamma^{*},Z^{*}\rightarrow H^{++}H^{--}$ and $q^{'}\overline{q}\rightarrow W^{*}\rightarrow H^{\pm\pm}H^{\mp}$ were studied  extensively in~\cite{Barger:1982, Gunion:1989, Muhlleitner:2003, Han:2007, Huitu:1997, Dion:1999, Akeroyd:2005}. Quartic terms in the scalar potential induce a mass splitting between $H^{\pm\pm}$ and $H^{\pm}$,  which can be of either sign. If $m_{H^{\pm\pm}} > m_{H^{\pm}}$ then a new decay channel becomes available for $H^{\pm\pm}$ , namely $H^{\pm\pm} \rightarrow H^{\pm} W^{*}$. 
Another scenario is the case of $m_{H^{\pm}} > m_{H^{\pm\pm}}$ , which would give rise to a new decay channel for the singly charged scalar, namely $H^{\pm} \rightarrow H^{\pm\pm} W^{*}$. This decay of singly charged Higgs would give rise to an alternative way to produce $H^{\pm\pm}$ in pairs, namely by the production mechanism $q^{'}\overline{q}\rightarrow W^{*}\rightarrow H^{\pm\pm}H^{\mp}$ followed by $H^{\mp} \rightarrow H^{\pm\pm} W^{*}$. This additional sources of pair produced $H^{\pm\pm}$ is expected to affect the loop induced Higgs diphoton decay, which I have studied in this article. It will be shown later  on that the contribution from $H^{\pm\pm}$ is four times compared to that from $H^{\pm}$ to the diphoton decay width. Thus the increase in the production of $H^{\pm\pm}$ is likely to enhance the decay width more rapidly as compared to the decrement arising from the decay of $H^{\pm}$. The contribution from $H^{\pm}$ is sub-dominant and thus increase in the production of $H^{\pm\pm}$ is likely to help us achieve the present LHC bounds on the diphoton decay width. I have made the same study in the 'Wrong Sign Limit' in  which some of the Yukawa couplings of the SM like Higgs boson are of the opposite sign to that of the vector boson couplings (wrong sign). The scalar spectrum of the model gets constrained by unitarity and the stability of the potential. The oblique T-parameter and Higgs decay branching ratios, in particular $h \rightarrow \gamma \gamma$ largely depend on the scalar spectrum of the model and thus constraints coming from their experimental values also restrict the scalar spectrum.

In section II, I have described the scalar potential and the physical mass eigenstates of the Higgs triplet model. Section III deals with the constraints on the parameter space of the model and section IV deals with the diphoton decay width. Additional channels for the production of doubly charged Higgs boson have been discussed in Section V followed by he numerical analysis of the effect of this additional production of $H^{\pm\pm}$ on $h\rightarrow\gamma\gamma$ in section VI. Section VII summarizes the article with the results and conclusion.

\section{The scalar potential and the physical mass eigenstates of the HTM}

As mentioned earlier, in a Higgs triplet model  a Y=2 complex $SU(2)_{L}$ isospin triplet of scalar fields, \textbf{T} =($T_{1}$ , $T_{2}$ , $T_{3}$ ), is added to the SM Lagrangian. Without the introduction of $SU(2)_{L}$ singlet neutrinos, Majorana masses can be obtained by the observed neutrinos in Higgs triplet model. The below gauge invariant Yukawa interaction accomplishes the task. Notation of~\cite{Akeroyd:2012}  has been followed here in describing the model. 

\begin{equation}
{\cal L}=h_{ll^{'}}L_{l}^{T}Ci\tau_{2}\Delta L_{l^{'}}+h.c.\,\,,
\end{equation}

where $h_{ll^{'}}$ ($l$,$l^{'}$ = $e$, $\mu$, $\tau$ ) is a complex and symmetric coupling, C is the Dirac charge conjugation operator, $\tau_{2}$ is the second Pauli matrix, $L_{l} = (\nu_{lL} , l_L)^{T}$ is a left-handed lepton doublet, and $\Delta$ is a 2$\times$2 representation of the Y = 2 complex triplet fields:

\begin{equation}
\Delta = \textbf{T}.\tau  = T_{1}\tau_{1} + T_{2}\tau_{2} + T_{3}\tau_{3} = \left(
\begin{array}{rcl}
\Delta^{+}/\sqrt{2}& \Delta^{++}\\
\Delta^{0}& -\Delta^{+}/\sqrt{2}
\end{array}
\right) \,\,,
\end{equation}

where $T_{1} = (\Delta^{++}+\Delta^{0})/2$, $T_{2} =i(\Delta^{++}-\Delta^{0})/2$, and $T_{3} = \Delta^{+}/\sqrt{2}$. Now $<\Delta^{0}>=\frac{v_{\Delta}}{\sqrt{2}}$ results in the following neutrino mass matrix: 
\begin{equation}
m_{ll^{'}}=2h_{ll^{'}}<\Delta^{0}>=\sqrt{2}h_{ll^{'}}v_{\Delta}
\,\,. \label{neutrino_mass}
\end{equation}

With the usual SM Higgs doublet defined as $\Phi=(\phi^{+}, \phi^{0})^{T}$, I move on to define the Higgs Triplet scalar potential~\cite{Ma:2000, Chun:2003}.

\begin{align}
V(\Phi,\Delta) &=
-m_{\Phi}^{2}\Phi^{\dagger}\Phi + \frac{\lambda}{4}(\Phi^{\dagger}\Phi)^{2}+ M_{\Delta}^{2}Tr\Delta^{\dagger}\Delta + (\mu\Phi^{T}i\tau_{2}\Delta^{\dagger}\Phi+h.c.)
\nonumber \\
 & \quad 
 + \lambda_{1}(\Phi^{\dagger}\Phi)Tr\Delta^{\dagger}\Delta + \lambda_{2}(Tr\Delta^{\dagger}\Delta)^{2} +\lambda_{3}Tr(\Delta^{\dagger}\Delta)^{2}+
\lambda_{4}\Phi^{\dagger}\Delta\Delta^{\dagger}\Phi\,\,.
\label{Potential}
\end{align}

$m_{\Phi}^{2}<0$ to ensure non-zero vev of the neutral component of the scalar doublet while $M_{\Delta}^{2}>0$. Here, $<\phi^{0}>=v/\sqrt{2}$ which spontaneously breaks the $SU(2)_L \otimes U(1)_Y$ to $U(1)_Q$. $v_{\Delta}$ is obtained from the  minimisation of V and for small $v_{\Delta}/v$ the expression for the triplet vev is,

\begin{equation}
v_{\Delta}\simeq \frac{\mu v^{2}}{\sqrt{2}(M_{\Delta}^{2}+v^{2}(\lambda_{1}+\lambda_{4})/2)}\,\,.
\end{equation}

For heavy triplet scalars, $M_{\Delta}\gg v$ and thus $v_{\Delta}$ can be approximated as $v_{\Delta}\simeq \mu v^{2}/ (\sqrt{2}M_{\Delta}^{2})$. As we can see, $v_{\Delta}$ will be naturally small even if $\mu$ is of the order of electroweak scale (this is sometimes called the "Type II seesaw mechanism"). Such heavy triplet scalars would be beyond the search limits of LHC. In the recent years there has been much interest in light triplet scalars($M_{\Delta}\approx v$) within the discovery reach of LHC. This would lead to $v_{\Delta}$ being approximately equal to $\mu$. It is to be noted that $v_{\Delta}$ has to be small basically for two reasons. First with reference to eq.(\ref{neutrino_mass}) where the neutrino mass matrix is directly proportional to $v_{\Delta}$ and thus to preserve the smallness of neutrino masses $v_{\Delta}$ has to be naturally small. Secondly the case of $v_{\Delta}<0.1$ MeV is assumed in the ongoing searches at the LHC, for which the BRs of the triplet scalars to leptonic final states (e.g. $H^{\pm\pm}\rightarrow l^{\pm}l^{\pm}$) would be $\sim$ 100\%. This fact has been discussed in~\cite{Akeroyd_Sugiyama:2011}. Since $v_{\Delta}\sim \mu$ for light triplet scalars then $\mu$ must also be small (compared to the electroweak scale) for the scenario of $v_{\Delta}<0.1$ MeV. 

 The $\rho$ parameter ($\rho=M_{W}^{2}/M_{Z}^{2}\cos^{2}\theta_{W}$) puts an upper bound on $v_{\Delta}$. In the SM $\rho=$ 1 at tree-level, while in the HTM it is,
 \begin{equation}
 \rho \equiv 1+ \delta \rho = \frac{1+2x^{2}}{1+4x^{2}}\,\,,
\end{equation}  
where $x=v_{\Delta}/v$. The measurement $\rho\approx 1$ leads to the bound $v_{\Delta}/v\lesssim 0.03$, or $v_{\Delta}\lesssim 8$GeV. Therefore the vev
of the doublet field $v$ is essentially equal to the vev of the Higgs boson of the SM (i.e. $v \approx$ 246 GeV).

Amongst the  physical mass eigenstates $H^{\pm\pm}$ is entirely composed of the triplet scalar field $\Delta^{\pm\pm}$ , while the remaining eigenstates are in general mixtures of the doublet and triplet fields. However, such mixing is proportional
to the triplet vev, and hence small even if $v_{\Delta}$ assumes its largest value of a few GeV. $\alpha$ is the mixing angle in the CP-even sector and $\beta^{'}$ is the mixing angle in the charged Higgs sector. Their expressions follow below.

\begin{equation}
\sin \alpha \sim 2v_{\Delta}/v\,\,\,,\,\,\, \tan \beta^{'}=\sqrt{2}v_{\Delta}/v\,\,.
\label{alpha_beta}
\end{equation}

The lighter CP-even Higgs, h is predominantly composed of the doublet field and plays the role of the SM Higgs boson. While the heavier CP-even Higgs, H, the singly charged Higgs, $H^{\pm}$ and the CP-odd Higgs, $A^{0}$ have predominant contribution from the triplet fields. 

Neglecting the small off-diagonal elements in the CP-even mass matrix, the approximate expressions for the squared masses of h and H are as follows:

\begin{equation}
m_{h}^{2}=\frac{\lambda}{2}v^{2}\,\,,
\label{mass_h}
\end{equation}

This is same as in the SM.

\begin{equation}
m_{H}^{2}= M_{\Delta}^{2}+(\frac{\lambda_{1}}{2}+\frac{\lambda_{4}}{2})v^{2}+3(\lambda_{2}+\lambda_{3})v_{\Delta}^{2}\,\,.
\label{mass_H}
\end{equation}

The squared mass of the doubly charged scalar $H^{\pm\pm}$ is,

\begin{equation}
m_{H^{\pm\pm}}^{2}= M_{\Delta}^{2}+\frac{\lambda_{1}}{2}v^{2}+\lambda_{2}v_{\Delta}^{2}\,\,.
\label{mass_H++}
\end{equation}

The squared mass of the singly charged scalar $H^{\pm}$ is,

\begin{equation}
m_{H^{\pm}}^{2}= M_{\Delta}^{2}+(\frac{\lambda_{1}}{2}+\frac{\lambda_{4}}{4})v^{2}+(\lambda_{2}+\sqrt{2}\lambda_{3})v_{\Delta}^{2}\,\,.
\label{mass_H+}
\end{equation}

The squared mass of the CP-odd scalar $A^{0}$ is,

\begin{equation}
m_{A^{0}}^{2}= M_{\Delta}^{2}+(\frac{\lambda_{1}}{2}+\frac{\lambda_{4}}{2})v^{2}+(\lambda_{2}+\lambda_{3})v_{\Delta}^{2}\,\,.
\label{mass_A}
\end{equation}

As we can figure out from the above equations that the mass of the SM like Higgs boson is unrelated to the rest of the masses and there is a common term in the expressions for the rest of the masses which is $M_{\Delta}^{2}+\frac{\lambda_{1}}{2}v^{2}$.
Neglecting the terms which are proportional to the small parameter $v_{\Delta}$, there are only two possible mass hierarchies for the non-standard scalars, with
the magnitude of the mass splitting being controlled by $\lambda_{4}$ (and $m_{A^{0}}=m_{H}$ when $v_{\Delta}$ is neglected):

\begin{eqnarray}
m_{H^{\pm\pm}}\,\,\,>\,\,\,m_{H^{\pm}}\,\,\,>\,\,\,m_{A^{0}},m_{H}\,\,\, for\,\,\,\lambda_{4}<0\,\,,\\\nonumber
m_{H^{\pm\pm}}\,\,\,<\,\,\,m_{H^{\pm}}\,\,\,<\,\,\,m_{A^{0}},m_{H}\,\,\, for\,\,\,\lambda_{4}>0\,\,.
\end{eqnarray}

\section{Constraints on the scalar potential}

The condition for the scalar potential in eq.(\ref{Potential}) to be bounded from below are:~\cite{Bonilla:2015} 

\begin{align}
\lambda\geq 0,\,\,\,\lambda_{2}+\lambda_{3}\geq 0,\,\,\,\lambda_{2}+\frac{\lambda_{3}}{2}\geq 0,\,\,\,\lambda_{1}+\sqrt{\lambda(\lambda_{2}+\lambda_{3})}\geq 0,\,\,\,\lambda_{1}+\lambda_{4}+\sqrt{\lambda(\lambda_{2}+\lambda_{3})}\geq 0\label{vacuum_stability}\\\nonumber
and\,\,\, \left[\vert\lambda_{4}\vert\sqrt{\lambda_{2}+\lambda_{3}}-\lambda_{3}\sqrt{\lambda}\geq0,\,\,\,or,\,\,\,2\lambda_{1}+\lambda_{4}+\sqrt{(2\lambda\lambda_{3}-\lambda_{4}^{2})(\frac{2\lambda_{2}}{\lambda_{3}}+1)} \geq 0 \right].
\end{align}
Theses are the conditions from vacuum stability. 
 
Scattering amplitudes involving longitudinal gauge bosons and Higgs bosons comprise the
elements of an S-matrix, having 2-particle states as rows and columns. The eigenvalues of this matrix are restricted by $\vert a_{0}\vert<1$, where $a_{0}$ is the l = 0 partial wave amplitude. These conditions translate into upper limits on combinations of Higgs quartic couplings, which for multi-Higgs models have been derived by different authors. For Higgs triplet model these have been derived in~\cite{Arhrib:2011} and are enlisted below:

\begin{eqnarray}
\vert(\lambda+4\lambda_{2}+8\lambda_{3})\pm \sqrt{(\lambda-4\lambda_{2}-8\lambda_{3})^{2}+16\lambda_{4}^{2}}\vert\leq 64\pi,\\
\vert(3\lambda+16\lambda_{2}+12\lambda_{3})\pm \sqrt{(3\lambda-16\lambda_{2}-12\lambda_{3})^{2}+24(2\lambda_{1}+\lambda_{4})^{2}}\vert\leq 64\pi,\\
\vert \lambda\vert\leq 32\pi,\\
\vert 2\lambda_{1}+3\lambda_{4}\vert \leq 32\pi,\\
\vert 2\lambda_{1}-\lambda_{4}\vert \leq 32\pi,\\
\vert \lambda_{1}\vert \leq 16\pi,\\
\vert \lambda_{1}+\lambda_{4}\vert \leq 16\pi,\\
\vert 2\lambda_{2}-\lambda_{3}\vert \leq 16\pi,\\
\vert \lambda_{2}\vert \leq 8\pi,\\
\vert \lambda_{2}+\lambda_{3}\vert \leq 8\pi.
\end{eqnarray}

Perturbativity constrains the quartic couplings to be within $[-4\pi,4\pi]$.

New Physics contribution to the electroweak T-parameter is given by~\cite{Chun:2012,Lavoura:1994},

\begin{equation}
\Delta T = \frac{1}{4\pi\sin^{2}\theta_{w}m_{W}^{2}}\left[F(m_{H^{\pm}}^{2},m_{A}^{2})
+F(m_{H^{\pm\pm}}^{2},m_{H^{\pm}}^{2})\right]\,\,,
\end{equation}
 where, $\theta_{w}$ is  the Weinberg angle and $m_{W}$ is the W-boson mass.
 The function $F(x,y)$ is defined as,
 \begin{equation}
 F(x,y)=\frac{x+y}{2}-\frac{xy}{x-y}\ln(\frac{x}{y})\,\,.
\end{equation}   

Experimentally the new Physics contribution to the T-parameter is given in~\cite{PDG:2014} to be $\Delta T <0.2$ at  $95\%$ C.L. This upper bound on $\Delta T$ translates into mass bounds on the non-standard scalar masses in HTM.  

The combined constraints from vacuum stability, unitarity, perturbativity and the electroweak T-parameter confine the parameter space for the masses of the non-standard scalars of the Higgs Triplet model. Since the masses of the non-standard Higgs bosons are correlated thus mass bound on one of these automatically puts a bound  on other masses too. These have been discussed  recently in~\cite{DDas:2016}.

\section{Diphoton decay in the Higgs Triplet model}
BSM Physics literature have shown the impact of singly charged scalars on the decay $h\rightarrow \gamma\gamma$ as for e.g. in the context of the minimal supersymmetric SM (MSSM)~\cite{Djouadi:1998}, a Two-Higgs Doublet Model~\cite{Ginzburg:2001,Arhrib:2004,Biswas:2016} and the Next-to-MSSM~\cite{Ellwanger:2012}. The contribution of doubly charged scalars to this decay has received comparatively very little attention. This was dealt in the Little Higgs Model~\cite{Han:2003}, but due to the theoretical structure of the scalar potential the magnitude of the contribution from $H^{\pm\pm}$ was shown to be much smaller than $H^{\pm}$. The contribution from $H^{\pm\pm}$ was studied in the HTM in~\cite{Kanemura:1201.6287,Arhrib:2012}, and was shown to give a sizeable contribution to $h\rightarrow \gamma\gamma$.

Now let me introduce the basic formula for diphoton decay width as written below~\cite{Shifman:1979},

\begin{align}
\Gamma(h\rightarrow \gamma \gamma) &=
\frac{G_{F}\alpha^{2}m_{h}^{3}}{128\sqrt{2}\pi^{3}}\mid\sum_{f}N_{c}Q_{f}^{2}g_{hff}A_{1/2}^{h}(\tau_{f})+g_{hWW}A_{1}^{h}(\tau_{W})
\nonumber \\
 & \quad 
 + \tilde{g}_{hH^{\pm}H^{\mp}}A_{0}^{h}(\tau_{H^{\pm}})+4\tilde{g}_{hH^{\pm\pm}H^{\mp\mp}}A_{0}^{h}(\tau_{H^{\pm\pm}})\mid^{2}\,\,.
\label{Decay_width}
\end{align}
 
 In the above equation, $\alpha$ is the fine structure constant, $N_{c}$ is the color quantum number which is 3 for quarks, $Q_{f}$ is the electric charge of the fermion in the loop and 
 \begin{equation}
  \tau_{i}=\frac{4m_{i}^{2}}{m_{h}^{2}},\,\,\,\,\,\, i= f,W,H^{\pm},H^{\pm\pm}\,\,.
  \label{tau_def}
  \end{equation}

The loop functions $A_{1/2}$ (for fermions), $A_{1}$ (for W-bosons) and $A_{0}$ (for the charged scalars) are defined below:
\begin{eqnarray}
A_{1/2}(\tau_{x})&=& -2\tau_{x}\left\{1+(1-\tau_{x}){\cal  F}(\tau_{x})\right\}\,\,,\\
A_{1}(\tau_{x})&=& 2+3\tau_{x}+3\tau_{x}(2-\tau_{x}){\cal  F}(\tau_{x})\,\,,\\
A_{0}(\tau_{x})&=& -\tau_{x}\left\{1-\tau_{x}{\cal  F}(\tau_{x})\right\}\,\,,\\ 
with,\,\,\,\,\,\, {\cal  F}(\tau_{x})&=& \left\{
\begin{array}{lr}
\left[\sin^{-1}(\sqrt{\frac{1}{\tau_{x}}})\right]^{2}&\,\,\,\,\,\,for\,\,\,\, \tau_{x}\geq 1,\\
-\frac{1}{4}\left[\ln (\frac{1+\sqrt{1-\tau_{x}}}{1-\sqrt{1-\tau_{x}}})-i\pi\right]^{2}&\,\,\,\,\,\, for\,\,\,\, \tau_{x}<1.
\end{array}
\right.
\end{eqnarray}
 
For the contribution from the fermion loops we will only keep the term with the top and bottom quarks, which are dominant.

There is an enhancement factor of four for $H^{\pm\pm}$ relative to  $H^{\pm}$. This is due to the electric charge. The couplings of h to the vector bosons and fermions relative to the values in the SM are as follows:
\begin{align}
g_{ht\overline{t}} &= \cos\alpha/\cos\beta^{'}\,\,,\label{coupling_htt}\\
g_{hb\overline{b}} &= \cos\alpha/\cos\beta^{'}\,\,,\label{coupling_hbb}\\
g_{hWW} &= \cos\alpha+2\sin\alpha v_{\Delta}/v\,\,,\label{coupling_hWW}\\
g_{hZZ} &= \cos\alpha+4\sin\alpha v_{\Delta}/v\,\,.\label{coupling_hZZ}
\end{align}

From eq.(\ref{alpha_beta}), it follows that $\cos\alpha= \sqrt{(1-4v_{\Delta}^{2}/v^{2})}\sim 1$ and $\cos\beta^{'}= \sqrt{(1-2v_{\Delta}^{2}/v^{2})}\sim 1$ and thus the above couplings of h are essentially the same as that of the SM Higgs boson because $v_{\Delta}\ll v$. This is the reason why h is called the SM-like Higgs boson.

The scalar trilinear couplings are parametrised as follows:
\begin{align}
\tilde{g}_{hH^{++}H^{--}} &= -\frac{m_{W}}{gm_{H^{\pm\pm}}^{2}}g_{hH^{++}H^{--}}\,\,,\label{hH++_coupling}\\
\tilde{g}_{hH^{+}H^{-}} &= -\frac{m_{W}}{gm_{H^{\pm}}^{2}}g_{hH^{+}H^{-}}\,\,.\label{hH+_coupling}
\end{align}

$g_{hH^{++}H^{--}}$ and $g_{hH^{+}H^{-}}$  are written below explicitly in terms of the parameters of the scalar potential (eq.\ref{Potential})~\cite{Arhrib:2011}:

\begin{align}
g_{hH^{++}H^{--}} &= -\left\{2\lambda_{2}v_{\Delta}s_{\alpha}+\lambda_{1}vc_{\alpha}\right\}\,\,,\label{H++Coupling}\\
g_{hH^{+}H^{-}} &= -\frac{1}{2}\left\{\left[4v_{\Delta}(\lambda_{2}+\lambda_{3})c_{\beta^{'}}^{2}+2v_{\Delta}\lambda_{1}s_{\beta^{'}}^{2}
-\sqrt{2}\lambda_{4}vc_{\beta^{'}}s_{\beta^{'}}\right]s_{\alpha}\right.\nonumber\\
 &+\left. \left[\lambda vs_{\beta^{'}}^{2}+(2\lambda_{1}+\lambda_{4})vc_{\beta^{'}}^{2}+
(4\mu-\sqrt{2}\lambda_{4}v_{\Delta})c_{\beta^{'}}s_{\beta^{'}}\right]c_{\alpha}\right\} \,\,,
\label{H+coupling}
\end{align}
where $s_{\alpha}=\sin\alpha$ and so on.

The above couplings take the simple forms when terms suppressed by $v_{\Delta}$ are neglected~\cite{Arhrib:2011,Akeroyd:2011}:
\begin{align}
g_{hH^{++}H^{--}}&\approx -\lambda_{1}v\,\,,\label{hH++_simplified_coupling}\\
g_{hH^{+}H^{-}}&\approx -(\lambda_{1}+\frac{\lambda_{4}}{2})v\,\,.\label{hH+_simplified_coupling}
\end{align}

The contribution from the $H^{\pm\pm}$ loop interferes constructively with that of the W-boson loop for $\lambda_{1} < 0$, while for $\lambda_{1} > 0$ the interference is destructive and its magnitude can be as large as that of the W contribution for $\lambda_{1}\sim 10$. The $H^{\pm}$ loop is usually sub-dominant. We also  note that it is essentially $\lambda_{1}$ which determines the value of $g_{hH^{++}H^{--}}$ and $g_{hH^{+}H^{-}}$. The main constraint on $\lambda_{1}$ comes from the requirement of the stability of the scalar potential, eq.(\ref{vacuum_stability}), and one of those constraints is,
\begin{equation}
\lambda_{1}+\sqrt{\lambda(\lambda_{2}+\lambda_{3})}\geq 0\,.
\label{constraint_lambda2_fixing}
\end{equation}
 
If $\lambda_{2}$ and $\lambda_{3}$ are taken to be zero, then the combined constraints on $\lambda_{1}$ from perturbative unitarity in scalar-scalar scattering and from stability of the potential require $\lambda_{1}>$0. However, if $\lambda_{2}$ and $\lambda_{3}$ are chosen to be sufficiently positive then negative values of $\lambda_{1}$ can be realized which is required for the calculation. Now the question is whether sufficiently positive values of $\lambda_{2}$ and $\lambda_{3}$ will affect the trilinear couplings and the masses of the triplet scalars or not? The answer is no. This is easily verifiable from eqs.(\ref{mass_H}\,\,\,-\,\,\,\ref{mass_A}\,\,\,,\,\,\,\ref{H++Coupling} and \ref{H+coupling}) where $\lambda_{2}$ and $\lambda_{3}$ come along with the sufficiently small parameter $v_{\Delta}$. In the numerical analysis $\lambda_{2}=\lambda_{3}$ have been fixed and eq.(\ref{constraint_lambda2_fixing}) has been used to determine $\lambda_{2}$ as a function of $\lambda_{1}$ and $\lambda$. $\lambda$ gets fixed from eq.(\ref{mass_h}) to be 0.516 ($m_{h}=125$ GeV and $v=  246$ GeV). Now choosing few specific values of $\lambda_{1}$, whose justification will be discussed soon, lower bounds on $\lambda_{2}$ and hence $\lambda_{3}$ can be calculated. For $\lambda_{1} = -1$, $\lambda = 0.516$ and using eq.(\ref{constraint_lambda2_fixing}), $\lambda_{2}\geq 0.97$ is obtained. Similarly for $\lambda_{1} = -2$, $\lambda = 0.516$ and using eq.(\ref{constraint_lambda2_fixing}), $\lambda_{2}\geq 3.9$ is obtained.

 Since further calculations have been done with $\lambda_{1}\,\, \epsilon \,\,[-2,0]$ hence the choice for $\lambda_{1}=-1$ and -2. As will be observed later from the graphical plots that for much negative values of $\lambda_{1}$ the diphoton decay width is much below the SM value. This is not the present LHC scenario and hence the choice for $\lambda_{1}$.










The relative decay width in Higgs triplet model is defined as,
\begin{equation}
\mu_{\gamma\gamma}=\frac{\sigma(pp\rightarrow h)^{HTM}\times\Gamma(h\rightarrow\gamma\gamma)^{HTM}}{\sigma(pp\rightarrow h)^{SM}\times\Gamma(h\rightarrow\gamma\gamma)^{SM}}\label{relative_decay_width}
\end{equation}

and the recent bounds on the relative decay width is $\mu_{\gamma\gamma}=1.16^{+0.20}_{-0.18}$~\cite{ATLASCMS:2015}.
 It is to be mentioned that since the lighter CP even Higgs boson, h of the Higgs Triplet model is considered to be the SM like Higgs boson so, its production cross-section from gluon gluon fusion is the same for HTM and SM.
 
\section{Additional sources of $H^{\pm\pm}$}

For $\lambda_{4}>0$, $m_{H^{\pm}}>m_{H^{\pm\pm}}$ thereby opening up the prospects of the decay channel $H^{\pm}\rightarrow H^{\pm\pm}W^{*}$. The decay rate after summing over all fermion states for $W^{*}\rightarrow f^{'}\overline{f}$, excluding the top quark is given below:

\begin{equation}
\Gamma(H^{\pm}\rightarrow H^{\pm\pm}W^{*}\rightarrow H^{\pm\pm}f^{'}\overline{f}) \simeq  \frac{9G_{F}^{2}m_{W}^{4}m_{H^{\pm}}}{4\pi^{3}}\int_{0}^{1-\kappa_{H^{\pm\pm}}}\,\,dx_{2}\int_{1-x_{2}-\kappa_{H^{\pm\pm}}}^{1-\frac{\kappa_{H^{\pm\pm}}}{1-x_{2}}}\,\,dx_{1}F_{H^{\pm\pm}W}(x_{1},x_{2})\,\,,
\label{H+_to_H++}
\end{equation}
where, $\kappa_{H^{\pm\pm}}\equiv m_{H^{\pm\pm}}/m_{H^{\pm}}$ and the analytical expression for $F_{ij}(x_{1},x_{2})$ is given by~\cite{Djouadi:1996,Moretti:1995},
\begin{equation}
F_{ij}(x_{1},x_{2})=\frac{(1-x_{1})(1-x_{2})-\kappa_{i}}{(1-x_{1}-x_{2}-\kappa_{i}+\kappa_{j})^{2}+\kappa_{j}\gamma_{j}}\,\,\,,
\end{equation}
with, $\gamma_{j}=\Gamma_{j}^{2}/m_{H^{\pm}}^{2}$.

This decay mode does not depend on $v_{\Delta}$. As long as the mass splitting between
$m_{H^{\pm}}$ and $m_{H^{\pm\pm}}$ is above the mass of the charmed hadrons ($\sim$ 2 GeV), f and $f^{'}$ can be taken to be massless to a good approximation. In my numerical
analysis I will be mostly concerned with sizeable mass splittings, $m_{H^{\pm}}-m_{H^{\pm\pm}}\gg 2$ GeV.

The other possible decays for $H^{\pm}$ are $H^{\pm}\rightarrow l^{\pm}\nu_{l^{'}}$, $H^{\pm}\rightarrow W^{\pm}Z$, $H^{\pm}\rightarrow W^{\pm}h$ and $H^{\pm}\rightarrow \overline{t}b$. The decay widths $\Gamma(H^{\pm}\rightarrow W^{\pm}Z)$, $\Gamma(H^{\pm}\rightarrow W^{\pm}h)$ and $\Gamma(H^{\pm}\rightarrow \overline{t}b)$ are greater than $\Gamma(H^{\pm}\rightarrow l^{\pm}\nu_{l^{'}})$ for $v_{\Delta}\gtrsim 0.1$ MeV while for $v_{\Delta}\lesssim 0.1$ MeV $\Gamma(H^{\pm}\rightarrow l^{\pm}\nu_{l^{'}})$ dominates. \cite{Akeroyd:2005, Chun:2003, Fileviez:2008, Chakrabarti_Gunion:1998} have already studied that the decay $H^{\pm\pm}\rightarrow H^{\pm}W^{*}$ can be the dominant decay channel for doubly charged scalar over a wide range of values of the mass difference between the doubly charged scalar and the singly charged scalar and $v_{\Delta}$. Similar result for the reverse decay of singly charged scalar to doubly charged scalar can be expected. The branching ratio BR($H^{\pm}\rightarrow H^{\pm\pm}W^{*}$) will be maximised with respect to $v_{\Delta}$ if $\Gamma(H^{\pm}\rightarrow l^{\pm}\nu_{l^{'}})= \Gamma(H^{\pm}\rightarrow W^{\pm}Z)+ \Gamma(H^{\pm}\rightarrow W^{\pm}h)+\Gamma(H^{\pm}\rightarrow \overline{t}b)$ which is realized for $v_{\Delta}\simeq$ 0.1 MeV. Since BR($H^{\pm}\rightarrow H^{\pm\pm}W^{*}$) needs to be maximised with respect to $v_{\Delta}$, thus I have worked with $v_{\Delta}=$ 0.1 MeV which is $\ll v(=246$ GeV) and thus all the approximations assumed above are inherently fulfilled.

From the production mechanism $q^{'}\overline{q}\rightarrow W^{*}\rightarrow H^{\pm\pm}H^{\mp}$ the decay mode, $H^{\pm}\rightarrow H^{\pm\pm}W^{*}$ would give rise to pair production of $H^{\pm\pm}$, with a cross section which can be comparable to that of the standard pair-production mechanism $q\overline{q}\rightarrow \gamma^{*}, Z^{*}\rightarrow H^{++}H^{--}$. This could significantly enhance the detection prospects of $H^{\pm\pm}$ in the four-lepton channel. In addition to the above, the decays $H\rightarrow H^{\pm}W^{*}$ and $A^{0}\rightarrow H^{\pm}W^{*}$ would also provide an additional source of $H^{\pm}$, which can subsequently decay to $H^{\pm\pm}$.

\section{Numerical analysis}
As has been discussed in the previous section that $H^{\pm\pm}$ has additional production channels above the standard one, thus this must affect the loop induced Higgs diphoton decay width. Now one may be inquisitive that since the additional production of $H^{\pm\pm}$ results from the decay of $H^{\pm}$ and thus the diphoton decay width will have reduced contribution from $H^{\pm}$ and enhanced contribution from $H^{\pm\pm}$. This can be reasoned out due to a factor of four in case of the contribution from $H^{\pm\pm}$ as compared to the contribution from $H^{\pm}$ in eq.(\ref{Decay_width}). Thus the enhancement is obviously higher that the decrement and this analysis has been made in the remaining part of this section. 

For the numerical analysis, ranges and values of the parameters required for the calculation have been chosen. Below they have been chalked out along with the justification for their choices.

\begin{itemize}
\item $\lambda_{4}>0$, for obtaining the mass hierarchy, $m_{H^{\pm\pm}}\,\,\,<\,\,\,m_{H^{\pm}}\,\,\,<\,\,\,m_{A^{0}},m_{H}$  which is in turn needed for the decay of $H^{\pm}$ to $H^{\pm\pm}$.
\item $\lambda_{1}<0$ for constructive interference between the combined SM contribution (from W boson and fermion loops) and contribution from $H^{\pm\pm}$.
\item $m_{H^{\pm}}=250$ GeV and $m_{H^{\pm\pm}}=200$ GeV since it has already been mentioned in~\cite{DDas:2016} that lower masses of $H^{\pm}$ and $H^{\pm\pm}$ give an enhancement to diphoton decay width w.r.t. SM as compared to heavier charged scalars. Moreover at lower masses, the non-standard scalars are not degenerate. Degeneracy encroaches at higher masses of the scalar particles. Degeneracy is unwanted here. Hierarchy plays the key role.
\item $v_{\Delta}=0.1$ MeV and thus the vev of the doublet, $v\sim 246$ GeV as constrained from the $\rho$ - parameter.
\end{itemize}

With reference to eq.(\ref{coupling_htt}\,\,-\,\,\ref{coupling_hWW}), $g_{ht\overline{t}}$, $g_{hb\overline{b}}$ and $g_{hWW}$ are $\sim 1$ for $v_{\Delta}=0.1$ MeV and $v\sim 246$ GeV as discussed. 
 



Now,  the  platter is ready to calculate the diphoton decay width.

When the lighter CP-even Higgs, h, in the HTM is the SM Higgs with $m_{h}=125$ GeV and $hV\overline{V}$ and $hq\overline{q}$ couplings are same as in the SM, then the diphoton decay width as mentioned in eq.(\ref{relative_decay_width}) can be  simplified as below:

\begin{equation}
 \mu_{\gamma\gamma}=\frac{\mid \frac{4}{3} A_{1/2}^{h}(\tau_{t})+\frac{1}{3} A_{1/2}^{h}(\tau_{b})+ A_{1}^{h}(\tau_{W})+\frac{(\lambda_{1}+\frac{\lambda_{4}}{2})}{2m_{H^{\pm}}^{2}}v^{2}A_{0}^{h}(\tau_{H^{\pm}})+4\times \frac{\lambda_{1}}{2m_{H^{\pm\pm}}^{2}}v^{2}A_{0}^{h}(\tau_{H^{\pm\pm}})\mid^{2}}{\mid \frac{4}{3} A_{1/2}^{h}(\tau_{t})+\frac{1}{3} A_{1/2}^{h}(\tau_{b})+ A_{1}^{h}(\tau_{W}) \mid^{2}}
\label{relative_decay_width_alignment_limit}
\end{equation}

Now let us define the \textit{Wrong Sign Limit} for Higgs triplet models. The wrong sign Yukawa coupling regime~\cite{Guedes:2014,Sampaio:2014,Ferreira:2014,Biswas:2016}
is defined as the region of 2HDM parameter space in which at least one of the couplings of the SM-like Higgs to up-type and down-type quarks is opposite in sign to the corresponding coupling of SM-like Higgs to vectors bosons.
This is to be contrasted with the Standard Model, where the couplings of $h_{SM}$ to $f\overline{f}$ and vector bosons are of the same sign.

The relative decay width in the Wrong Sign Limit (WSL) takes the below form:

\begin{equation}
\mu_{\gamma\gamma}^{WSL}=\frac{\mid \frac{4}{3} A_{1/2}^{h}(\tau_{t})-\frac{1}{3} A_{1/2}^{h}(\tau_{b})+ A_{1}^{h}(\tau_{W})+\frac{(\lambda_{1}+\frac{\lambda_{4}}{2})}{2m_{H^{\pm}}^{2}}v^{2}A_{0}^{h}(\tau_{H^{\pm}})+4\times \frac{\lambda_{1}}{2m_{H^{\pm\pm}}^{2}}v^{2}A_{0}^{h}(\tau_{H^{\pm\pm}})\mid^{2}}{\mid \frac{4}{3} A_{1/2}^{h}(\tau_{t})+\frac{1}{3} A_{1/2}^{h}(\tau_{b})+ A_{1}^{h}(\tau_{W}) \mid^{2}}
\label{relative_decay_width_WSL}
\end{equation}

I have plotted $\mu_{\gamma\gamma}$ vs $\lambda_{4}$ and $\mu_{\gamma\gamma}^{WSL}$ vs $\lambda_{4}$ for various values of $\lambda_{1}$. The quartic couplings $\lambda_{1}$ and $\lambda_{4}$ are responsible for $H^{\pm}$ to $H^{\pm\pm}$ decay and constructive contribution from $H^{\pm\pm}$ loop. 

\begin{figure}[h]
\begin{center}
\begin{minipage}[h]{0.45\linewidth}
\includegraphics[width=\linewidth]{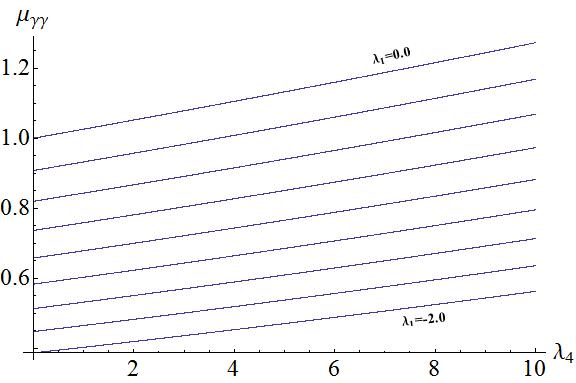}
(a)
\end{minipage}
\hspace{0.5cm}
\begin{minipage}[h]{0.45\linewidth}
\includegraphics[width=\linewidth]{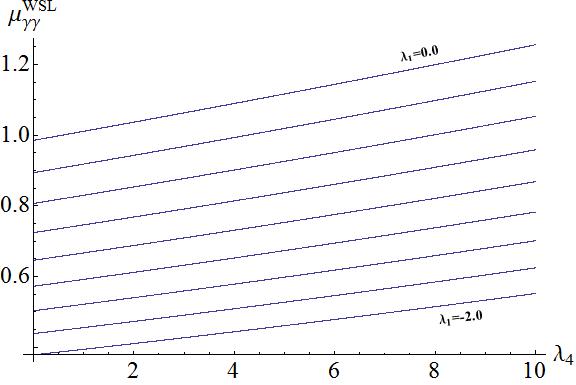}
(b)
\end{minipage}
\end{center}
\caption{\small Diphoton decay width of the SM-like Higgs particle (normalized to SM) 
		as a function of $\lambda_{4}$ for $m_{H^{\pm}}=250$GeV, $m_{H^{\pm\pm}}=200$GeV and various negative values of $\lambda_{1}$\,, for (a) same sign and (b) wrong sign, of
		down-type Yukawa couplings in Higgs Triplet model. }
		\label{diphoton.fig} 
\end{figure}
  
\section{Results and Conclusion}
From fig.(\ref{diphoton.fig}), a few observations are made. Firstly, for any particular value of $\lambda_{1}$, $\mu_{\gamma\gamma}$ and $\mu_{\gamma\gamma}^{WSL}$ increases as $\lambda_{4}$ increases. This indicates that as the mass difference between $H^{\pm}$ and $H^{\pm\pm}$  increases ($m_{H^{\pm}}^{2}-m_{H^{\pm\pm}}^{2}=\frac{\lambda_{4}}{4}v^{2}$), relative diphoton decay width increases since the tendency for $H^{\pm}$ to decay to $H^{\pm\pm}$ increases. More the number of $H^{\pm\pm}$ produced more the increment since there is an enhancement factor of 4 which comes from the electric charge in case of $H^{\pm\pm}$'s contribution to $h\rightarrow\gamma\gamma$ as compared to that of $H^{\pm}$. Moreover higher value of $\mid \lambda_{1}\mid$ is not required to obtain the experimentally  obtained enhancement in the relative diphoton decay width. Thus by maintaining a mass hierarchy in the parameter space possibilities for the singly charged scalar to decay to doubly charged scalar have been generated and by choosing a particular range of a quartic coupling, $\lambda_{1}$, which controls the contribution of these non-standard charged scalars to the diphoton decay width an enhancement in the diphoton decay width relative to the SM value has been obtained that lies well within the experimental bounds.

\begin{figure}[htbp]
\includegraphics[height=0.5\columnwidth, width = 0.6\columnwidth]{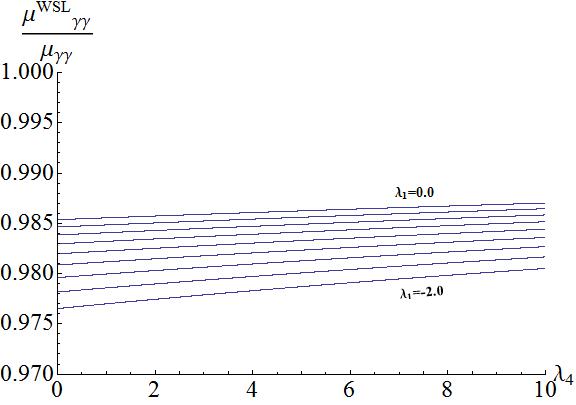}
\caption{$h\gamma\gamma$ decay width for `wrong sign' $h\bar{D}D$ coupling relative to the
	case with `same sign' Yukawa couplings}
\label{relative.fig}
\end{figure}

The ratio of the diphoton decay width when the down type Yukawa coupling have \textit{wrong sign} relative to the case with the \textit{same sign} Yukawa coupling has been plotted in fig.(\ref{relative.fig}).  We can easily point out that this ratio varies within a very narrow range and converges for higher values of $\lambda_{4}$.

In passing, I comment that although I have assumed $v_{\Delta}=0.1$ MeV, the above conclusions do not crucially depend on the numerical value of $v_{\Delta}$ as long as it remains small.

I conclude by saying that if tighter bounds are imposed on the discovery of $H^{\pm\pm}$ and $H^{\pm}$ via the four lepton(4\textit{l}) and three lepton(3\textit{l}) channels and their mass ranges be confined, then these non-standard scalars may account for the excess in the loop induced Higgs to diphoton decay. In that case the decay of $H^{\pm}$ to $H^{\pm\pm}$  must also be taken into account for the enhancement. In future more precise search for charged scalars may open up prospects for Higgs Triplet Model and further constrain the parameter space of the Higgs Triplet Model when experimental bounds are superimposed on the theoretical predictions. This will in turn signal that the Higgs discovered at the LHC is a part of a richer scalar sector and not merely the Standard Model.

\begin{acknowledgements}
I thank A.~Lahiri for many helpful discussions during different stages of this work.
\end{acknowledgements}

      
\end{document}